\documentclass[conference]{IEEEtran}
\IEEEoverridecommandlockouts
\usepackage{cite}
\usepackage{amsmath,amssymb,amsfonts}
\usepackage{algorithmic}
\usepackage{graphicx}
\usepackage{comment}
\usepackage{textcomp}
\usepackage{amsthm}
\usepackage{xcolor}
\usepackage{bbding}
\usepackage{makecell}
\usepackage{algorithm}  
\usepackage{color}
\usepackage{multirow}
\usepackage{bm}
\usepackage{array}
\usepackage{booktabs}
\usepackage{tabularx,booktabs}
\usepackage{multicol}
\usepackage{cleveref}
\usepackage{balance}
\usepackage{lastpage}
\usepackage{tabulary}
\usepackage{subfigure}
\usepackage{etoolbox}
\usepackage{bbm}
\usepackage{enumitem}
\usepackage{mathrsfs}
\usepackage{multicol}
\usepackage{amsfonts,amssymb}
\usepackage{setspace}
\usepackage{ulem}
\usepackage{stackengine}
\usepackage{fancyhdr} % 添加页眉页脚
\usepackage{threeparttable}

\usepackage{amsthm}
\theoremstyle{definition}

\def\BibTeX{{\rm B\kern-.05em{\sc i\kern-.025em b}\kern-.08em
    T\kern-.1667em\lower.7ex\hbox{E}\kern-.125emX}}

\begin{document}

\bstctlcite{IEEEexample:BSTcontrol}

\title{An Overview of Machine Learning-Enabled Optimization for Reconfigurable Intelligent Surfaces-Aided 6G Networks: From Reinforcement Learning to Large Language Models\\
%\thanks{%Identify applicable funding agency here. If none, delete this.}
}

\author{\IEEEauthorblockN{Hao Zhou, \IEEEmembership{Member, IEEE}, Chengming Hu, \IEEEmembership{Student Member, IEEE}, Xue Liu, \IEEEmembership{Fellow, IEEE}.}
\IEEEauthorblockA{\textit{School of Computer Science, McGill University, Montreal, Quebec, Canada}}
\{hao.zhou4, chengming.hu\}@mail.mcgill.ca, xueliu@cs.mcgill.ca 
}

\maketitle

\thispagestyle{fancy}            %更改plain状态，首页格式设为fancy
\chead{This is an invited work towards IEEE Future Networks World Forum. } 

\renewcommand{\headrulewidth}{1pt}      %把页眉线的宽度设为零，即去掉页眉线
\pagestyle{plain}

\begin{abstract}
Reconfigurable intelligent surface (RIS) becomes a promising technique for 6G networks by reshaping signal propagation in smart radio environments.
However, it also leads to significant complexity for network management due to the large number of elements and dedicated phase-shift optimization.
In this work, we provide an overview of machine learning (ML)-enabled optimization for RIS-aided 6G networks. In particular, we focus on various reinforcement learning (RL) techniques, e.g., deep Q-learning, multi-agent reinforcement learning, transfer reinforcement learning, hierarchical reinforcement learning, and offline reinforcement learning.
Different from existing studies, this work further discusses how large language models (LLMs) can be combined with RL to handle network optimization problems. It shows that LLM offers new opportunities to enhance the capabilities of RL algorithms in terms of generalization, reward function design, multi-modal information processing, etc. 
Finally, we identify the future challenges and directions of ML-enabled optimization for RIS-aided 6G networks.
\end{abstract}


\section{Introduction}
Reconfigurable intelligent surface (RIS) is a promising technology for 6G networks, providing attractive properties to reshape the signal propagation path \cite{yliu}. 
Many studies have demonstrated that RISs can significantly improve the network channel capacity, signal coverage, transmission security, energy efficiency, etc \cite{zhou2023survey}. 
In addition, RISs also have low energy consumption and hardware cost, and
therefore they can be easily deployed in various scenarios, such as building walls and ceilings, to improve the signal transmission environment.
Given the great potential, RISs have been combined with many other state-of-the-art techniques, such as mmWave and THz communications, non-terrestrial networks, unmanned aerial vehicles (UAVs), vehicle-to-everything (V2X) networks, non-orthogonal multiple access (NOMA), and so on.
Despite the improved performance, it is worth noting that integrating RISs into existing wireless networks may considerably increase the network optimization complexity \cite{zhou2023survey}. 
In particular, RISs may consist of hundreds of small elements, and each unit requires dedicated control of the phase shifts and on/off status, leading to large solution spaces with non-convex objectives and constraints.
Convex optimization is a widely considered approach for RIS-related optimization problems, e.g., alternating optimization, majorization-minimization algorithm, successive convex optimization, semidefinite relaxation, and others.  
However, these optimization problems are usually highly non-convex and non-linear, which must be reformulated to convex formats.
Moreover, there may be integer control variables due to resource allocation and associate problems, and the mixed integer non-linear programming (MINLP) problem is a well-known challenge for convex optimization techniques.

To this end, machine learning (ML) offers promising opportunities to overcome these complexities \cite{kfai}. For instance, reinforcement learning (RL) is the most widely used ML-enabled optimization algorithm, and many optimization problems can be easily transformed into unified Markov decision processes (MDPs) \cite{zhou2021ran}. In particular, the optimization objectives, such as achieved data rates and coverage, are defined as rewards, control variables like RIS phase shifts become actions, and network dynamics are transformed into environment states, e.g., channel state information and previous data rates. 
In this way, RL algorithms can be applied to try different action combinations under various states, aiming to maximize the long-term network performance. 
It indicates a unified optimization scheme regardless of the problem convexity or continuity, lowering the difficulty of optimizing network performance. 
Meanwhile, RL can also be combined with other ML techniques, indicating diverse RL algorithms with unique advantages. 
For instance, integrating transfer learning and RL can produce transfer reinforcement learning (TRL), enabling knowledge sharing between previously completed tasks and incoming new tasks \cite{zhou2022knowledge}. 
Such a knowledge reuse scheme can be very useful for solving a series of tasks with similarities, which are very common in wireless networks.

Recent years have witnessed the rapid progress of ML techniques, and specifically large language models (LLMs) have received considerable interest from academia and industry, leading to revolutionary changes in many fields \cite{zhao2023survey}. 
Compared with conventional ML techniques, LLMs have demonstrated outstanding comprehension and reasoning capabilities after the pre-training on large datasets, e.g., instruction following, in-context learning, and zero-shot learning, among others~\cite{brown2020language}.
For example, in-context learning means that LLMs can learn from contextual inputs such as task descriptions and demonstrations, and then apply these hidden patterns to the downstream tasks directly. 
By contrast, in previous ML algorithms, learning a new policy usually requires dedicated design, indicating iterative and time-consuming model training or fine-tuning.
Furthermore, LLMs can utilize contextual feedback from previous tasks to improve the implementation of the next task, which is known as self-reflection. Recent studies reveal that LLMs can work as agents to improve the performance of target tasks by learning from the environment feedback iteratively \cite{shinn2023reflexion}. 
%
%Therefore, it means that LLMs can be used to solve optimization problems with proper task description and feedback design. 
%
Such optimization capabilities become more promising when combined with LLM's instruction-following features. For instance, operators can use natural language to instruct the network management policy, and then the LLM-based agent can improve the overall policy automatically. 

Given the above advantages of LLM techniques, it is crucial to investigate the applications of state-of-the-art ML techniques to optimize RIS-aided 6G networks, paving the way to artificial general intelligence (AGI)-enabled 6G.    
In this work, we provide a comprehensive overview of using ML to optimize RIS-aided 6G networks, ranging from various reinforcement learning algorithms to LLM-aided optimization techniques.  
Specifically, we present an in-depth discussion on state-of-the-art LLM techniques for optimization problems, which is different than many existing studies on RIS-related optimization problems.
The rest of this work is organized as the following. Section \ref{sec-related} introduces related studies, and Section \ref{sec-overview} overviews the problem formulations and features of RIS-related optimization problems. Section \ref{sec-rl} presents various RL algorithms for RIS-related optimization tasks, and Section \ref{sec-llm} discusses LLM-aided optimization techniques. Finally, Section \ref{sec-future} identifies challenges and future directions and Section \ref{sec-con} concludes this work.

\section{Related Works}
\label{sec-related}
There have been many studies that apply ML techniques to optimize RIS-aided wireless networks. 
For instance, Faisal and Choi summarized various ML algorithms for RISs in \cite{kfai}, including supervised learning, unsupervised learning, reinforcement learning, federated learning, etc.
Zhou \textit{et al.} presented a comprehensive survey on model-based, heuristic, and ML algorithms for RIS-aided wireless networks in \cite{zhou2023survey}, and they further explored heuristic-aided ML for RIS optimization in \cite{zhou2023heuristic}. 
Meanwhile, Liu \textit{et al.} introduced RIS beamforming, resource management and ML algorithms for RIS-aided networks in \cite{yliu}.
Puspitasari and Less provided an overview of RISs and RL technique implementations to optimize RIS design and operation \cite{puspitasari2023survey}.
The above studies \cite{kfai,yliu,puspitasari2023survey,zhou2023heuristic,zhou2023survey} have investigated various ML algorithms for RIS-aided wireless networks. 
Especially, RL is involved in most of the previous studies due to its critical importance in handling optimization problems, e.g., the RL framework proposed in \cite{yliu}, heuristic DQN in \cite{zhou2023heuristic}, and transfer RL in \cite{zhou2023survey}.

However, the ML field is advancing rapidly, and LLM is considered the most state-of-the-art ML technique, leading to revolutionary changes in many other fields such as education, finance and healthcare. 
As a subfield of generative artificial intelligence (GenAI), researchers have started exploring LLM applications in wireless networks. 
For instance, Lin \textit{et al.} investigated the LLM deployment at the network edge in \cite{lin2023pushing}, and overviewed several key techniques for efficient edge model training and inference. 
Xu \textit{et al.} discussed some topics to integrate LLMs into wireless networks, including end-edge-cloud collaboration, integrated sensing and communication, and digital twin techniques\cite{xu2024large}.
Similarly, Shen \textit{et al.} also introduced the LLM-enabled edge AI, indicating that LLMs in the central cloud can process the requests from network edge servers\cite{shen2024large}.
These useful studies have demonstrated the great potential of LLM-empowered wireless networks.

This work is different from existing studies \cite{kfai,yliu,puspitasari2023survey,zhou2023heuristic,zhou2023survey} by introducing a comprehensive overview of various RL algorithms and discussions on LLM-aided optimization techniques for RIS-aided 6G networks. 
In particular, we discuss how LLMs can be combined with RL schemes to improve the generalization capabilities, reward function design, multi-task handling, etc. 
LLM-empowered RL techniques offer new opportunities for more efficient and accessible optimization technologies for 6G networks.

\begin{figure}[!t]
\centering
\includegraphics[width=0.9\linewidth]{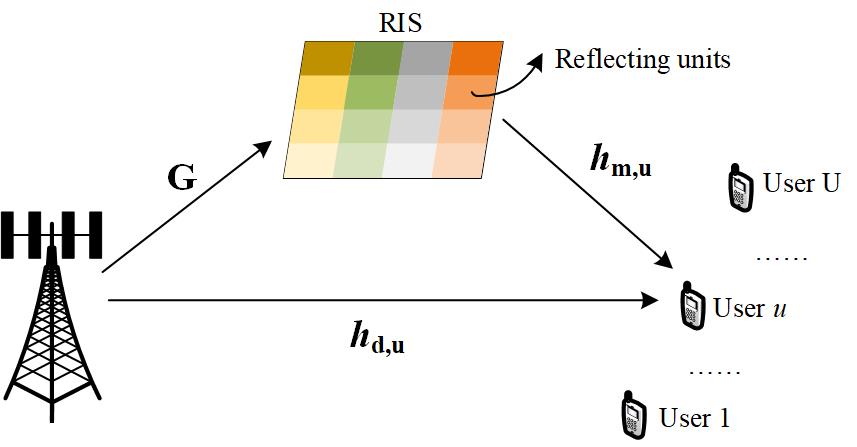}
\caption{Downlink transmission system for RIS-aided networks.}
\label{fig-ris}
\setlength{\abovecaptionskip}{-2pt} 
\end{figure}

\section{An Overview of RIS-related Optimization Problems}
\label{sec-overview}
This section will first overview the RIS-related optimization problems in terms of problem formulations, and then we will analyze the problem convexity and complexity.
Fig. \ref{fig-ris} presents a RIS-aided downlink transmission system with $N$ RIS reflecting units, one base station (BS) with $M$ antennas, and $U$ single-antenna users. 
The RIS units provide an extra transmission path, indicating that users can receive signals by two links, direct link BS-user and RIS-aided link BS-RIS-user. 
With RISs, the signal-to-interference-plus-noise ratio (SINR) of user $u$ in a multiple input single output (MISO) system is:
\begin{equation}\label{eq7}
\Gamma_{u}=\frac{|(\bm{h}^{RIS}_{u}\bm{\Theta}\bm{G}+\bm{h}^{D}_{u})p_{u}|^2}{\sum\limits_{j=1,j\neq u}^{U}|(\bm{h}^{RIS}_{u}\bm{\Theta}\bm{G}+\bm{h}^{D}_{u})p_{j}|^2+N_{0}^2},
\end{equation}
where $h^{D}_{u} \in \mathbb{C}^{1 \times M}$ represents the channel gain of a direct link from the BS to the user $u$, and the RIS-aided link is indicated by $\bm{h}^{RIS}_{u}\bm{\Theta}\bm{G}$. In particular, $h^{RIS}_{u} \in \mathbb{C}^{1 \times N}$ indicates the channel gain from RIS elements to user $u$, $\bm{G}\in \mathbb{C}^{N \times M}$ represents the channel gain from BS antennas to RIS elements, and RIS units are defined by $\bm{\Theta}=\text{diag}(\theta_{1},\theta_{2},...,\theta_{n},...,\theta_{N})\in \mathbb{C}^{N\times N}$. 
Here each RIS unit has a specific phase-shift vector $ \theta_{n}=e^{j\phi_{n}}$. Finally, $p_{u}$ and $p_{j}$ are the BS antenna transmit power for users, and $N_{0}^2$ shows the transmission noise.

Equation (\ref{eq7}) reveals that the phase shifts of RIS elements will affect the SINR $\Gamma_{u}$ of users.
After that, various problem formulations can be defined, e.g., maximizing sum rate/channel capacity, signal coverage, energy efficiency, network fairness, and secrecy rate, or minimizing the power consumption, which has been investigated in many existing studies \cite{zhou2023survey}.  
For example, sum-rate maximization problem is usually formulated as:
\begin{subequations}\label{e-sum}
\begin{align}
\max\limits_{p,\bm{\Theta}}  \qquad & \sum_{u=1}^{U}w_{u} \log(1+\Gamma_{u})  & \tag{\ref{e-sum}} \\
 \text{s.t.}  \qquad & \sum_{u=1}^{U}||p_{u}||^2\leq P_{max}, & \label{e-sum1}\\
  \qquad &|\theta_{n}|=1, n=1,2,...,N,  & \label{e-sum2}
\end{align}
\end{subequations}
where $w_{u}$ is the pre-defined weight of user $u$, and $P_{max}$ is the BS antenna's maximum transmission power. 
The objective in equation (\ref{e-sum}) is to maximize the sum-rate of all users under the maximum transmission power constraint (\ref{e-sum1}) and RIS operation constraint (\ref{e-sum2}).  
The optimization problem defined in equation (\ref{e-sum}) is obviously highly non-convex and non-linear. Specifically, it first involves multiple logarithm terms, and each item includes a fractional term $\Gamma_{u}$. Meanwhile, the constraint (\ref{e-sum2}) is related to the 2-norm of value $\theta_{n}$. 
In practice, the phase shift of RIS elements may be constrained by discrete values with $\theta_n \in \{0, \frac{2\pi}{2^\varrho},...,(2^\varrho-1)\frac{2\pi}{2^\varrho},2\pi\}$, where $\varrho$ is the resolution of RIS elements. This is a more practical setting by using discrete phase-shifts to replace the continuous RIS phase-shift control. But it will also lead to MINLP problems, which are very hard to solve efficiently by using conventional convex optimization techniques.       
On the other hand, the control variables $p$ and $\bm{\Theta}$ mean a large solution space, since the RISs may consist of hundreds of small units.
Moreover, RIS technology is usually integrated into other techniques, such as NOMA, V2X, UAV, etc. These combinations will introduce other control variables, e.g., user decoding order in NOMA and UAV position and power control, which will further increase the difficulties of finding optimal solutions.

In summary, the above examples and analyses demonstrate that optimizing RIS-related problems can be extremely difficult, especially in dynamic wireless environments.
To this end, ML techniques, especially RL, offer promising solutions to improve RIS-aided network performance, which will be introduced in the following section.

\begin{figure*}[!t]
\centering
\includegraphics[width=1\linewidth]{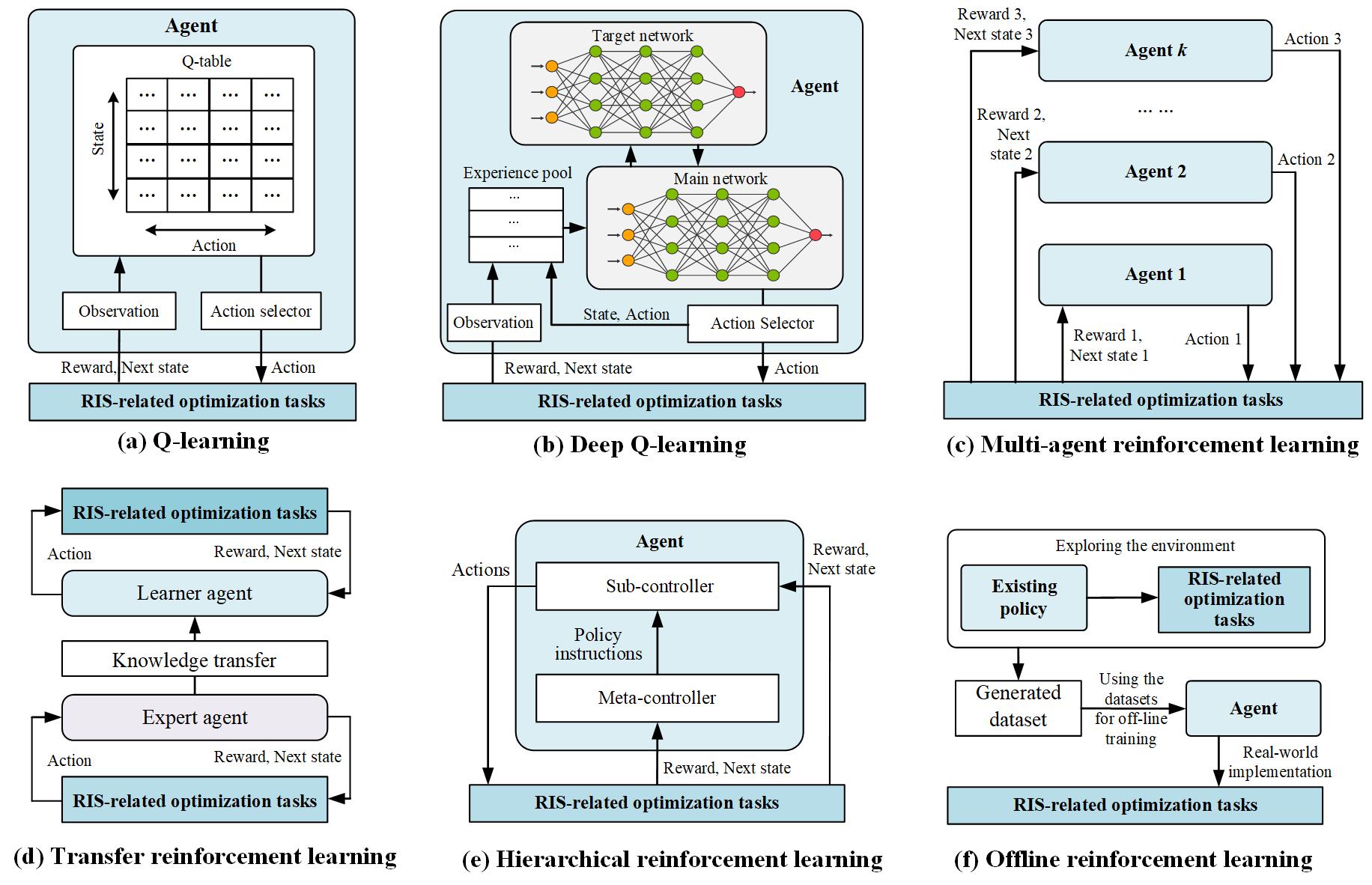}
\caption{Overview of various reinforcement learning techniques.}
\label{fig-rl}
\setlength{\abovecaptionskip}{-2pt} 
\end{figure*}

\section{Reinforcement Learning-based Optimization}
\label{sec-rl}
This section will introduce various RL algorithms to solve RIS-related optimization problems. It first presents RL fundamentals, especially MDP definitions, and then discusses various RL techniques, including Q-learning, deep Q-learning (DQN) and variants, multi-agent reinforcement learning (MARL), TRL, hierarchical reinforcement learning (HRL), and offline reinforcement learning (offline-RL).

\subsection{Reinforcement Learning Fundamentals}
Defining MDPs is the prerequisite for implementing RL algorithms, which consist of a tuple $<s, a, r, T>$ with states $s$, actions $a$, rewards $r$, and transition probability $T$.
In particular, one agent can select one action $a$ based on the current state $s$, receive an instant reward $r$ from the environment, and then move to the next state $s'$. The agent aims to maximize the long-term accumulated reward by trying different action combinations and balancing the exploration and exploitation policies.
Proper MDP definition is crucial for using RL, since the states, actions, and rewards are closely related to the objectives, control variables, and constraints in RIS-related optimization problems.
\begin{itemize}
    \item \textbf{State definitions}: The state should reflect the environment status, and then the agent can make decisions accordingly. For instance, it is obvious that RIS phase-shift control should consider the channel condition, and then channel state information is frequently defined as the state in many existing studies\cite{zhou2023survey}. 
    Other than that, current RIS unit phases and positions, energy level, and previous data rate can also contribute to state definitions. 
    
    \item \textbf{Action definitions}: The actions in an MDP are usually equivalent to the control variables in optimization problems. 
    For example, RIS phase shifts and BS transmission power in the equation (\ref{e-sum}) should be considered as actions, since these decisions will affect the reward. 
    The action space may extend when other control variables are involved, e.g., RIS location optimization and elements on/off control.       
    
    \item \textbf{Reward definitions}: The reward definition is closely related to the optimization objective. Maximizing the long-term reward indicates improving the corresponding objective function iteratively. Therefore, the reward in RIS-related optimization problems usually depends on specific objectives, e.g., sum-rate, energy efficiency, secrecy rate, signal coverage, etc.    
\end{itemize}
With the above MDP definitions, a standard optimization problem as shown in equation (\ref{e-sum}) can be transformed into a unified scheme. Then various RL algorithms can be applied.

\subsection{From Q-learning to Deep Q-learning and Variants }

Q-learning is the most fundamental RL algorithm that has been extensively investigated, which employs a Q-table to record all the state-action values.
However, such a tabular-based approach cannot handle problems with large state-action spaces, indicating many iterations to converge the state-action values.
To this end, DQN is proposed by using neural networks to predict the state-action values. Two useful techniques in DQN are experience relay and double networks. 
In particular, as illustrated in Fig.\ref{fig-rl} (b), experience reply means collecting previous experience in terms of MDP tuples $<s,a,r,s'>$, which will be stored in the experience pool and then sampled for neural network training.
Meanwhile, double networks involve the main network to predict the current state-action values, and the target network to provide a stable target state-action value for main network training. 
Based on the DQN framework, many other variants have been proposed to further improve the algorithm performance, including deep actor-critic, deep deterministic policy gradient (DDPG), double deep Q-learning (DDQN), twin delayed DDPG (TD3), etc. 
For instance, DDPG can handle problems with continuous action spaces, which is particularly useful for RIS phase-shift optimization problems for continuous phase control.

\subsection{Multi-agent Reinforcement Learning (MARL)}
Multi-agent problems are very common in RIS-aided wireless networks. For instance, RISs with different geographical locations can be considered as different agents, and each agent has its own objectives, e.g., maximizing the sum-rate of a microcell or minimizing the transmission power consumption of the serviced area. 
In particular, a multi-agent system indicates that each agent can make decisions autonomously to improve its objective, which depends on the environment and also the actions of other agents\cite{zhou2021ran}.  
Therefore, the core of a multi-agent system is to coordinate the policies of each local agent, aiming to optimize the overall performance of the whole system.
A straightforward coordination approach is to define a unified reward function to represent the overall system performance, and then all local agents will select actions to prioritize the performance of the whole system, e.g., maximize the data rate of the whole macrocell.
Other than that, nash equilibrium is also a useful method to solve MARL problems, which means that no agent has anything to gain by changing only one agent's policies.
In addition, other techniques such as correlated equilibrium and matching theory, can also be combined with MARL to solve multi-agent problems in RIS-aided wireless networks.

\begin{table*}[!t]
\caption{Summary of RL techniques for RIS-aided 6G networks. }
\centering
\small
\setstretch{1.1}
\resizebox{1\textwidth}{!}{%
\begin{tabular}{|m{2cm}<{\centering}|m{7cm}<{\centering}|m{5cm}<{\centering}|m{5cm}<{\centering}|}
\hline \makecell{RL \\ techniques} &  \makecell{Main features \\ and advantages} & \makecell{Potential \\ difficulties} & \makecell{RIS-aided 6G network \\ applications}\\
\hline
DQN and variants & DQN provided the fundamental framework for many deep reinforcement learning techniques, which is also the most widely used RL techniques. There have been many variants based on the DQN framework, e.g., deep actor-critic, deep deterministic policy gradient (DDPG), double deep Q-learning, and twin delayed DDPG. Each method has its unique advantages such as handling continuous action spaces and overcoming over-estimation.      & A common problem of conventional DQN is the long training time, which means many iterations to produce a stable model. Meanwhile, another well-known problem is the hyperparameter selection and fine-tuning, e.g., neural network learning rate and number of hidden layers.   &   DDPG is particularly useful for solving RIS phase-shift optimization problems, since the action space is continuous for phase shift control.  Meanwhile, DQN can also be used for discrete RIS phase-shift optimization problems if $\theta_n \in \{0, \frac{2\pi}{2^\varrho},...,(2^\varrho-1)\frac{2\pi}{2^\varrho},2\pi\}$. \\
\hline
Multi-agent reinforcement learning & MARL focuses on problems with multiple elements and various objectives. In particular, each agent should be able to make decisions autonomously, aiming to improve its objective by interacting with the environment and other agents. It provides a reasonable approach to modelling complicated relationships between multiple agents.    &  The main difficulty lies in the coordination mechanism design, since the action of one agent will affect both the environment and other agents. The corresponding techniques include unified reward function, nash equilibrium, correlated equilibrium, etc.     &   MARL can be applied to RIS-aided wireless networks when multiple network elements are involved, e.g., multiple RISs, network slices, and unmanned aerial vehicles (UAVs). MARL can jointly consider the service requirements of these elements for coordination.\\
\hline
Transfer reinforcement learning &  Transfer learning indicates an efficient approach to process a series of similar tasks. By reusing previous knowledge on former tasks, transfer learning enables shorter training iterations and faster convergence than conventional RL.         & The main difficulty of transfer reinforcement learning lies in the definition of mapping functions. The reason is that knowledge may exist in various forms in the RL scheme, and the mapping function must transform existing knowledge to make it accessible for learner agents.  & 
Transfer learning aligns well with the inherent wireless communication features to enable faster response time. Based on existing experience in RIS phase-shift optimization, transfer learning can be applied to more complicated scenarios in which RISs are combined with other techniques, such as RIS-UAVs and RIS-NOMA.          \\
\hline
Hierarchical reinforcement learning &  HRL introduces hierarchical architecture to conventional RL schemes, allowing for hierarchical decision-making by multiple control layers. It usually includes one meta-controller to produce high-level policies, and then lower-level sub-controllers can make specific decisions based on environment states and high-level policies.  &   The hierarchy architecture increases system flexibility, but it also leads to concerns about the system stabilities. In particular, the main difficulty is how to guarantee the quality of the long-term policies. This is crucial since it will further affect the actions of the sub-controller.  &     HRL allows for multiple decision variables with different time scales, which is very common in RIS-aided wireless networks. For instance, the power consumption level is usually a long-term performance metric, while latency and sum-rate represent instant metrics. HRL scheme may be used to jointly optimize these indicators.            \\
\hline
Offline reinforcement learning &  Offline RL offers a practical and realistic technique to train RL agents without interfering with the real environment.  It uses offline datasets to train the RL agent, overcoming the cost and risks of training a new agent from scratch in real-world scenarios.   &  The performance of offline RL algorithm is closely related to the dataset quality. To be specific, the RL agent can learn good decision-making strategies from datasets generated by high-quality policies. By contrast, randomly generated datasets may lead to poor learning performance.  &  Offline-RL is a promising technique for RIS-aided 6G networks. It enables offline model training and learning without touching the real wireless network environment, which is much more practical than many existing online algorithms.                       \\
\hline
\end{tabular}}
\label{tab-summary}
\end{table*}

\subsection{Transfer Reinforcement Learning (TRL)}
Transfer learning provides an efficient approach to reuse the experience on previous tasks, and then transfers the knowledge to current target tasks\cite{zhou2022learning}. 
Specifically, as shown in Fig.\ref{fig-rl} (d), the decision-making of conventional RL $\mathcal{D}_{RL}$ can be defined as $\mathcal{D}_{RL}:s \times \mathcal{K}\stackrel{a}{\Longrightarrow} <r, s'>$. It means that the agent selects actions $a$ under current state $s$ by using its knowledge $\mathcal{K}$, and then receives a reward $r$ and moves to the next state $s'$.  
By contrast, TRL decision-making $\mathcal{D}_{TRL}$ is: $\mathcal{D}_{TRL}:s \times \mathcal{M}(\mathcal{K}_{expert}) \times \mathcal{K} \stackrel{a}{\Longrightarrow} <r, s'>$. 
Compared with conventional RL, the main difference lies in $\mathcal{M}(\mathcal{K}_{expert})$. $\mathcal{K}_{expert}$ indicates the existing knowledge that is achieved by previous tasks of the expert agent. 
However, it is worth noting that $\mathcal{K}_{expert}$ may exist in various forms, e.g., neural network weights, state-action values, or RIS phase-shift control decisions. Therefore, a mapping function $\mathcal{M}$ is needed to transform the expert knowledge into a more accessible format for the learner agent.
TRL can be a useful technique for RIS-aided 6G networks to shorten the algorithm training, which is considered one of the bottlenecks of applying ML techniques to real-world applications.
In addition, it also enables faster response time to network dynamics, and better adapts to changing wireless environments.

\subsection{Hierarchical Reinforcement Learning (HRL)}
HRL introduces hierarchies into the standard RL agent. As illustrated in Fig.\ref{fig-rl} (e), HRL includes a meta-controller and a sub-controller. 
Both controllers can receive rewards and states from the environment, while the meta-controller can send policy instructions to the sub-controller. 
To be specific, the MDP of the meta-controller is $<s,p,r,T>$, and here we use $p$ to replace the action $a$ in conventional MDP, which indicates the policies made by the meta-controller based on current state $s$ \cite{zhou2023cooperative}. Then, the MDP of the sub-controller is $<s,p,a,r,T>$, which means that the action is selected based on both current state $s$ and policy $p$ from the meta-controller: $\mathcal{D}_{RL}:s \times p \times \mathcal{K}\stackrel{a}{\Longrightarrow} <r, s'>$. 
HRL can be used to solve problems with multiple decision variables that have different timescales.
For instance, Zhou \textit{et al.} applied HRL to combine BS sleep control with RIS phase-shift optimization, in which the meta-controller decides the BS on/off status to save long-term energy consumption, and the sub-controller optimizes the RIS phase shifts to improve the data rate \cite{zhou2022hierarchical}.

\subsection{Offline Reinforcement Learning (Offline RL)}
Conventional RL schemes improve policy performance through interactions with the environment to collect rewards and feedback~\cite {levine2020offline}.
However, such an online approach may be impractical for many real-world scenarios, e.g., autonomous driving or healthcare applications.
In contrast, offline RL trains the RL agent using offline datasets without direct interactions with the real-world environment. This approach can be more realistic and practical for real-world applications, due to the advantages of learning from pre-collected data without the limitations of real-time constraints. 
For instance, training an RL agent directly in realistic RIS-aided 6G networks is impractical due to the critical importance of telecommunication services.
Building a specific simulator for the training can also be time-consuming. 
To this end, offline RL provides a promising solution by training the agent using existing datasets collected from the environment.
However, it is worth noting that the performance and effectiveness of offline RL  depend on the quality of offline datasets used for training, as the agent cannot access the realistic wireless environment. 
Hence, ensuring that these datasets are comprehensive and representative of various scenarios is crucial for achieving robust performance of offline RL.

\section{LLM-enabled Optimization Techniques}
\label{sec-llm}

Although RL algorithms can solve a range of RIS-related optimization problems, these techniques commonly encounter some potential challenges, such as difficulties in generalizing to unseen environments. 
TRL and offline RL techniques can improve generalization to a certain extent, but they still require significant model training to adapt to new environments. 
Moreover, designing reward functions in RL typically necessitates specialized domain knowledge for target tasks, and conventional trial-and-error design methods can be time-consuming. 
To address these limitations, LLMs have shown several promising features, which may be combined with RL algorithms:

\begin{itemize}
    \item \textbf{Instruction following}: 
    One core capability of LLMs is to follow natural language instructions, which improves the generalization ability across a wide range of language-related tasks~\cite{wei2021finetuned}. 
    The common format of natural language instructions includes a task description, an optional input, a corresponding output, and several examples as demonstrations. 
    By constructing instructions in this manner, LLMs can be fine-tuned using supervised sequence-to-sequence training loss, which aids in enhancing language comprehension across multiple languages.  
    Hence, LLMs can follow a broader range of instructions, thereby enhancing their adaptability and generalization in multi-lingual contexts.
    
    \item \textbf{Multi-modal learning}: 
    Multi-modal learning is a notable capability of LLMs, enabling the model to process related information from multiple modalities, including text, audio, image, 3D maps, etc~\cite {zhang2023meta}. 
    A multi-modal LLM can employ various encoders to extract features from multi-modal inputs into desired outputs, providing a broader perspective and greater adaptability when processing multi-source information.
    
    \item \textbf{In-context learning}: 
    In-context learning utilizes formatted natural language prompts along with task descriptions to guide LLMs in recognizing and performing new tasks based on understanding contextual information~\cite{brown2020language}. 
    Specifically, LLMs identify the task type by interpreting the provided prompts and task descriptions, drawing on pre-existing knowledge from their pre-training data~\cite{wies2024learnability}.
    After task identification, LLMs employ the included demonstrations to learn solution strategies, enabling efficient implementation of new tasks.
    
    \item \textbf{Step-by-step reasoning}: 
    LLMs exhibit strong performance in complex multi-step tasks through advanced prompting strategies, including chain-of-thought (CoT) prompting, tree-of-thought (ToT) prompting, graph-of-thought (GoT) prompting, etc.  
    These strategies organize the problem-solving process into sequential or hierarchical steps, providing a more structured and comprehensive reasoning pathway. 
    For instance, CoT prompting augments demonstrations into a sequence of reasoning steps~\cite{wei2022chain}, which allows LLMs to solve complex problems with greater transparency and logic. 
\end{itemize}

The above capabilities indicate great potential for integrating LLMs into RL schemes, including better generalization capabilities, automated reward design, multi-modality information processing, multi-task handling, and LLM-aided reasoning. In the following, we will introduce the potential benefits in detail. 
\begin{itemize}
    \item \textbf{Improving generalization capabilities:} LLMs have been pre-trained by a large number of real-world datasets, which means that the system model can handle many realistic problems in a zero-shot manner.
    LLM's real-world knowledge may complement the RL's generalization capabilities to better adapt to unseen environments.
    For instance, LLM may provide prior knowledge for RL agents in a transfer learning manner, and then the algorithm can quickly understand new environments, achieving a jump-start with a very small number of iterations.
    Such generalization improvement is crucial for the real-world applications of RL algorithms, especially considering the complicated RIS-aided wireless environments.
    
   \item  \textbf{Automated reward function design}: In addition, with proper prompt designs, existing studies have shown that LLMs can be used to generate reward functions for RL algorithms, achieving comparable performance with manually designed rewards\cite{ma2023eureka}. 
   For example, wireless network performance may involve multiple indicators, such as latency, throughput, and packet drop rate, and balancing these metrics manually in a comprehensive reward design can be difficult. 
   With proper instructions, LLM can automate the reward function design, balancing multiple network indicators based on human preference and system feedback.
   Such a technique can significantly save human efforts on reward function design, enabling automated RL schemes for RIS-aided 6G.

   \item \textbf{Multi-modality information processing for RL}:
   Multi-modality refers to the capability of processing related information with different modalities, such as text, image, audio, video, and graphs. 
   Combining multi-modal LLMs with RL algorithms will improve RL agents’ comprehension of complicated scenarios with diverse information sources. 
   For instance, integrated sensing and communication has become an important technique for future 6G networks, and LLM-aided RL can make the most of sensing information, e.g., audio, images, and 3D maps, to better capture environment dynamics.
   Such comprehensive multi-modal input will improve the decision-making quality of RL agents, e.g., using environment images as input for RIS-aided beamforming.

   \item \textbf{Multi-task handling}: Conventional RL algorithms usually implement one single task, while LLMs can easily handle multiple downstream tasks simultaneously using billions of parameters.
   It means one single LLM can serve multiple RL agents with different purposes, enabling efficient RL model training and fine-tuning on diverse target tasks.
   Multi-task handling can improve the RL system efficiency, processing a series of tasks efficiently with the assistance of LLMs, e.g., RIS-aided resource allocation, RIS elements on/off control and elements phase-shift optimization, etc.   

   \item \textbf{LLM-enabled reasoning for RL}
    Despite the great performance in handling optimization problems, existing RL techniques lack reasoning capabilities. In particular, it means that the RL agent cannot provide reasons behind the selected decisions. For instance, after dedicated model training, the RL algorithm may decide to decrease the BS transmission power to achieve higher energy efficiency in RIS-aided wireless networks.
    However, the agent is unable to provide the intellectual oversight behind this decision, since decreasing the transmission power may also lower the data rate. 
    Integrating LLMs into the RL scheme may provide explanations for the optimal policies selected by RL algorithms, and humans can better understand the insight of optimizing RIS-aided wireless networks.     
\end{itemize}

\begin{figure}[!t]
\centering
\includegraphics[width=0.95\linewidth]{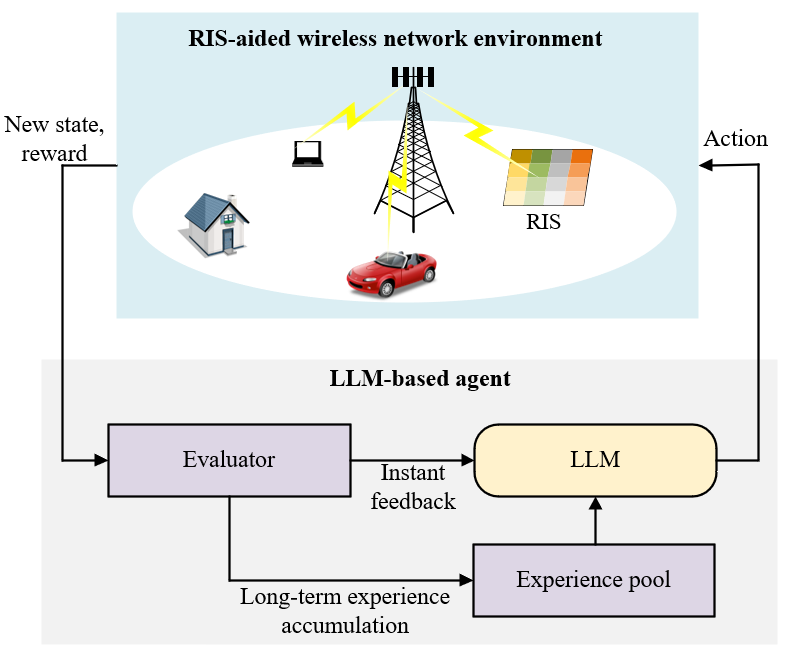}
\caption{LLM as an agent for RIS-aided networks.}
\label{fig-llmrl}
\end{figure}

Finally, Fig.\ref{fig-llmrl} presents a scheme of using LLM as an agent to solve optimization problems in RIS-aided wireless networks.
The LLM will first select an action, and receive new states and rewards from the RIS-aided wireless network environment.
Then, the evaluator can provide instant feedback to LLM to evaluate the previous actions, e.g., "\textit{The last selected action is bad because the data rate is too low, and you need to provide another solution to achieve a higher data rate}". 
On the other hand, the evaluator should send some possible experience to the experience pool, which will accumulate long-term experience for the LLM agent reference, e.g., "\textit{Users with low-latency requirements should be prioritized in the service queue}". Combining instant feedback with long-term experience shows a similar design as an instant reward and experience pool in DQN algorithms.  

\section{Challenges and Future Directions}
\label{sec-future}

This section will discuss the challenges and future directions of ML-enabled optimization for RIS-aided 6G networks, including practical RIS optimization techniques, wireless domain-specific LLM training, and LLM's demand for computational resources.

\begin{itemize}
    \item \textbf{Practical RIS optimization techniques:} A large number of techniques have been proposed to optimize RIS-aided wireless networks, e.g., convex optimization, heuristic algorithms, and ML methods. However, most existing techniques rely on ideal assumptions for the wireless environment, e.g., static and evenly distributed users, perfect channel state information acquisition, and instant information sharing and processing. 
    In real-world applications, RIS control may face more difficult and complicated environments, e.g., limited computational resources, information-sharing delays between RISs and BSs, imperfect channel state information acquisition, etc.
    Solving these problems will further facilitate the real-world applications of RIS technologies. 
    
    \item \textbf{Wireless domain-specific LLM training:} Previous sections have discussed the great potential of integrating LLM-aided RL into RIS-aided wireless networks. 
    However, some wireless-specific problems may require professional knowledge and understanding of network architecture, channel modelling, signal processing, and so on.
    Most existing LLM models are designed for general-domain purposes, lacking wireless domain-specific knowledge.
    Therefore, a practical approach is to fine-tune a general-domain LLM such ChatGPT, LLaMA, and Google Gemini, to adapt to wireless service requests.
    Such a wireless-LLM may bring revolutionary changes to the operation of RIS-aided 6G networks.
    
    \item \textbf{LLMs are computationally demanding}: 
    Although previous sections have discussed the advantages of LLM-enabled optimization techniques, there remain significant challenges in deploying LLMs into RIS-aided 6G networks due to intensive computational resource requirements.  
    For instance, LLM inference time can significantly contribute to system latency, with response time ranging from 0.58 to 90 seconds~\cite{9512383}, which can be a major obstacle to latency-critical applications. 
    Furthermore, the complex architectures of LLMs generally require extensive computational requirements, leading to a misalignment with the limited computational resources and storage capacity of network devices.  
    Hence, These challenges present considerable obstacles to the scalability and feasibility of deploying LLMs in resource-constrained environments, highlighting the need for further exploration to deploy and optimize LLMs into real-world 6G networks.
\end{itemize}

\section{Conclusion}
\label{sec-con}

This work provides a comprehensive overview of various RL techniques to solve RIS-related optimization problems, including Q-learning, DQN and variants, MARL, TRL, HRL, and offline RL.
LLMs have demonstrated remarkable capabilities in many real-world applications, and this paper further explores the potential of LLM-enabled optimization techniques by integrating LLMs with various RL techniques, aiming to enhance the performance in RIS-aided 6G networks. 
Finally, the potential challenges of LLM-enabled optimization techniques suggest future directions in several areas, including developing advanced RIS optimization techniques, creating wireless domain-specific LLMs, and deploying computationally efficient LLMs.

%\section*{Acknowledgment}

\normalem
\bibliographystyle{IEEEtran}
\bibliography{Reference}

\end{document}